\documentclass{article}
\usepackage{ijcai19}

\usepackage{times}
\usepackage{soul}
\usepackage[hidelinks]{hyperref}
\usepackage[utf8]{inputenc}
\usepackage[small]{caption}
\usepackage{graphicx}
\usepackage{amsmath}
\usepackage{amsfonts}
\usepackage{booktabs}

\usepackage{algorithm}
\usepackage{algpseudocode}
\newcommand\Algphase[1]{%
\vspace*{-0.5\baselineskip}\Statex\hspace*{\dimexpr-\algorithmicindent-2pt\relax}\rule{8.56cm}{0.4pt}%
\Statex\hspace*{-\algorithmicindent}\textbf{#1}%
\vspace*{-0.5\baselineskip}\Statex\hspace*{\dimexpr-\algorithmicindent-2pt\relax}\rule{8.56cm}{0.4pt}%
}
\newcommand{\squeezeup}{\vspace{-2.3mm}}
\newcommand{\squeezeups}{\vspace{-1.0mm}}

\title{Real-Time Adversarial Attacks}

\author{
Yuan Gong
\and
Boyang Li \and
Christian Poellabauer \And
Yiyu Shi 
\affiliations
University of Notre Dame\\
\emails
\{ygong1, bli1, cpoellab, yshi4\}@nd.edu
}

\begin{document}

\maketitle

\begin{abstract}
In recent years, many efforts have demonstrated that modern machine learning algorithms are vulnerable to adversarial attacks, where small, but carefully crafted, perturbations on the input can make them fail. While these attack methods are very effective, they only focus on scenarios where the target model takes \textbf{static} input, i.e., an attacker can observe the entire original sample and then add a perturbation at any point of the sample. These attack approaches are not applicable to situations where the target model takes \textbf{streaming} input, i.e., an attacker is only able to observe past data points and add perturbations to the remaining (unobserved) data points of the input. In this paper, we propose a \textbf{real-time adversarial attack} scheme for machine learning models with streaming inputs.

\end{abstract}

\section{Introduction}
Over the last decade, machine learning has made great advances and has been widely adopted for many diverse applications, including security-sensitive applications such as identity verification and fraud detection. However, recent research has also shown that many machine learning algorithms (specifically deep neural networks) are vulnerable to \emph{adversarial attacks}, where small, but carefully designed, perturbations are added to original samples, leading the target model to make wrong predictions~\cite{szegedy2013intriguing}. Such adversarial attack algorithms have been proposed for a variety of tasks, such as image recognition, speech processing, text classification, and malware detection, where they have also been shown to be highly effective~\cite{moosavi2016deepfool,cisse2017houdini,gong2017crafting,alzantot2018did,carlini2018audio,schonherr2018adversarial,ebrahimi2017hotflip,grosse2017adversarial}.

Most existing adversarial example generation algorithms require that the \textbf{entire} original data sample that is fed into the target model is observed and that \textbf{any part} of the sample can then be modified. For example, speech adversarial attack algorithms typically design a perturbation for a given speech sample, add the perturbation to the original sample, and then feed the resulting sample into the target speech recognition system. However, this approach is not always feasible, particularly when the target system requires {\em streaming input}, where the input is continuously processed as it arrives. In this real-time processing scenario, an attacker can only observe \textbf{past parts} of the data sample and can only add perturbations to \textbf{future parts} of the data sample, while the decision of the target model will be based on the entire data sample. A few concrete scenarios that operate this way are as follows:

\begin{figure}[t]
  \centering
  \includegraphics[width=7.5cm]{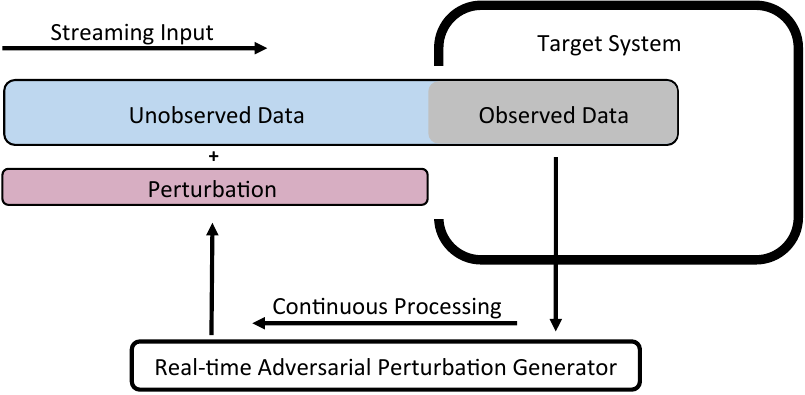}
  \caption{An illustration of the real-time adversarial attack scheme. The target system takes streaming input; only past data points can be observed and adversarial perturbation can only be added to future data points. The adversarial perturbation generator continuously uses observed data to approximate an optimal adversarial perturbation for future data points.}
  \label{fig:system_ilus}
  
\end{figure}

\textbf{Financial Trading Systems.}
Financial institutions make trading decisions using automatic machine learning algorithms based on a sequence of observations of some market conditions (e.g., variations in the stock index). An attacker may influence the trading model's outcomes by carefully perturbing the corresponding market conditions. However, while the target trading model usually makes decisions based on a long sequence of observations, the attacker cannot change any historical data. Instead, the attack can only add perturbations to future (yet to be observed) market conditions, e.g., using market manipulations.

\textbf{Real-time Speech Processing Systems.}
Machine learning based real-time speech processing systems (e.g., speech recognition and automatic translation systems) have been adopted widely, including many security-sensitive applications. An attacker may want to change the output of such systems by playing a carefully designed noise that is unnoticeable by the human ear, but will be superimposed on the speech generated by a human speaker through the air. The attacker can only design such noise signals based on past speech signals and superimpose the noise only on future speech signals, while the speech processing system will perform its task using the entire speech segment (e.g., a word or a sentence).
    
When attacking a real-time system, the attacker faces a trade-off between \emph{observation} and \emph{action space}. That is, assume that the target system takes a sequential input $\boldsymbol{x}$, the attacker could choose to design adversarial perturbations at the beginning. However, in this case, the attacker does not have any observation of $\boldsymbol{x}$, but perturbations can be added to any time point of $\boldsymbol{x}$, i.e., the attacker has {\em minimum observation} and {\em maximum action space}. In contrast, if the attacker chooses to add adversarial perturbations at the end, the attacker has a full observation of $\boldsymbol{x}$, but cannot add perturbations to the data (i.e., the attacker has {\em maximum observation}, but {\em minimum action space}). In the first case, it is hard to find an optimal perturbation for $\boldsymbol{x}$ without having any observations, while in the second case, the attack cannot be implemented at all. To address this dilemma, we propose a new attack scheme that continuously uses observed data to approximate an optimal adversarial perturbation for future time points using a deep reinforcement learning architecture (illustrated in Figure~\ref{fig:system_ilus}). In this paper, we refer to such attacks as \textbf{real-time adversarial attacks}. To the best of our knowledge, this is the first study of dynamic real-time adversarial attacks, which have not yet received the attention they deserve. The closest related concept is \emph{universal adversarial perturbation}, presented in~\cite{moosavi2017universal,li2018adversarial,neekhara2019universal}, where the authors design a fixed adversarial perturbation that is effective for different samples. The main difference to our work is that the universal adversarial perturbation is built offline and does not take advantage of observations in real-time to further improve the perturbation for a specific target input.

\section{Real-time Adversarial Attacks}

\subsection{Problem Formalization}
\label{sec:formalism}

Let $\boldsymbol{x} = \{x_1, x_2, x_3, ..., x_n \}\in\mathbb{R}^{m\times n}$ denote an $n$-point time-series data sample,  where each point $x_i\in\mathbb{R}^m$; $f: \mathbb{R}^{m\times n} \longrightarrow \{1 \dots k\}$ is a classifier mapping the time-series sample $\boldsymbol{x}$ to a discrete label set. The goal of the attacker is to design a real-time adversarial perturbation generator $g(\cdot)$ that continuously uses observed data $\{x_1, x_2, ..., x_t\}$ to approximate an optimal adversarial perturbation $r_{t+d+1}$ for a future time point $t+d+1$, where $d$ is the delay caused by processing the data or emitting the adversarial perturbation. That is,

\begin{equation}
\label{equ:1}
r_{t} = 
\begin{cases}
    g(\{x_1, x_2, ..., x_{t-d-1}\}) \quad \text{$d+1 <t \leq n$} \\
    $0$ \quad \text{else}
\end{cases}
\end{equation}

We define a metric $m(\cdot)$ to measure the perceptibility of the adversarial perturbation; a common choice for $m$ is the induced metric of $l_p \, (p\in\{0,1,2,\inf\})$ norm. We then aim to solve the following optimization problem for non-targeted adversarial attacks:

\begin{equation}
\label{equ:2}
\begin{aligned}
\text{minimize} &\quad m(\boldsymbol{r}=\{r_1, r_2, ..., r_n\}) \\
\quad \text{s.t.} &\quad f(\boldsymbol{x} + \boldsymbol{r})\neq f({\boldsymbol{x}})
\end{aligned}
\end{equation}

Equation \ref{equ:1} implies the constraint that adversarial perturbation is crafted only based on the observed part of the data sample and can only be applied to the unobserved part of the data sample. Equation \ref{equ:2} implies that the attacker wants to make the perturbation as imperceptible as possible on the premise that the attack succeeds. Even without the constraint of Equation \ref{equ:1}, directly solving Equation \ref{equ:2} is usually intractable when $f$ is a deep neural network due to its non-convexity. Nevertheless, previous efforts have found effective approximation methods such as the fast gradient sign method (FGSM)~\cite{goodfellow2014explaining}, DeepFool~\cite{moosavi2016deepfool}, and the algorithm proposed in~\cite{carlini2018audio}. However, all these methods require full observation and the freedom of changing any point of the original data sample, and therefore these methods are not compatible with the constraint imposed by Equation \ref{equ:1}. 

Alternatively, a more natural way of describing this problem is to view the adversarial perturbation generator as an \emph{agent} and model the problem as a \emph{partially observable decision process} problem, i.e., the generator continuously observes the streaming data and makes a sequence of decisions of how to make the perturbation. This formalism is equivalent to Equations \ref{equ:1} and \ref{equ:2}, but allows us to use the many tools available for reinforcement learning (RL)~\cite{sutton1998introduction} to solve the problem. Then, the problem can be described using a tuple $\langle O, S, A, T, R\rangle$, where:

\begin{enumerate}
    \item \textbf{Observation $O$:} $o_t=\{x_1, x_2, ..., x_t\}$.
    
    \item \textbf{State $S$:} unobservable hidden state.
    
    \item \textbf{Action $A$:} $a_t=r_{t+d+1}$, i.e., adding the perturbation to the original sample at time $t+d+1$.
    
    \item \textbf{Transition $T$:} unknown.
    
    \item \textbf{Reward $R$:} $\boldsymbol{I}_{f(\boldsymbol{x} + \boldsymbol{r})\neq f({\boldsymbol{x}})} - m(\boldsymbol{r})$. 
\end{enumerate}

This means that the attacker performs an action $a_t$ to emit the perturbation valued $r_{t+d+1}$ at $t+d+1$ based on the observation $o_t$, which will change the internal hidden state according to an unknown transition rule (e.g., the state can be the attack success probability, and an action could make it increase or decrease). The adversarial generator will only get the reward at the end. The goal of RL is to learn an optimal policy $\pi_g: a_{t}=g(o_t)$ that maximizes the expectation of the reward. In this problem, the environment is the target model $f$, and the input data distribution $P_{\boldsymbol{x}}$.

\subsection{Adversarial Attacks Using Reinforcement Learning}

As discussed in the previous section, real-time adversarial attacks can be described as reinforcement learning problems, which are usually solved by using deep neural networks (DNNs). RL-DNN based adversarial attacks and conventional optimization based adversarial attacks (e.g., FGSM and DeepFool) differ in that the former treats the original example and the corresponding adversarial perturbation as the input and output of an unknown nonlinear mapping and then use a DNN to approximate it, i.e., \textbf{use learning to substitute optimization}. In geometric terms, the attack model is trying to predict the direction that pushes the original example $\boldsymbol{x}$ out of the correct decision region using the shortest distance.

\begin{figure}[t]
  \centering
  \includegraphics[width=4.3cm]{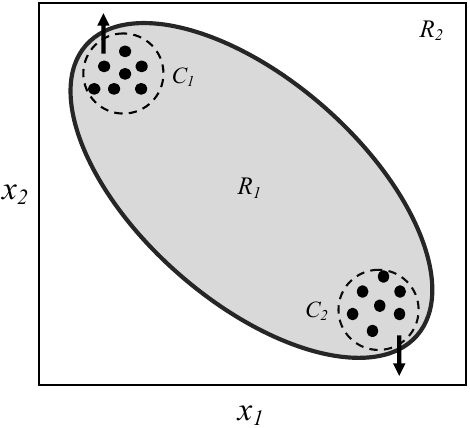}
  \caption{A low dimensional illustration of how the attack model predicts perturbations for future data points based on already observed data points. The attacker is trying to push the original data samples (shown as solid dots) from the decision region $\mathcal R_1$ to $\mathcal R_2$. Without observation of $x_1$, the attacker indeed has no idea about the optimal perturbation direction on $x_2$, but after observing $x_1$, the attacker knows the optimal perturbation direction on $x_2$, i.e., up if $\boldsymbol{x}$ is in cluster $C_1$ and down if $\boldsymbol{x}$ is in cluster $C_2$.}
  \label{fig:ilus}
  \squeezeup
  \squeezeup
\end{figure}

A challenge for the attack model is to ``forecast'' future perturbations on yet unobserved data. However, this is feasible since, given a specific machine learning task, the input sample, although yet unobserved, will obey some fixed distribution (e.g., distribution of natural speech), and there usually exist dependencies among the data points of the data sample, which can be used to forecast some characteristics of future data points based on already observed data points. We expect that such characteristics contain information that can be used to estimate an optimal perturbation for future points, which is illustrated in Figure~\ref{fig:ilus}.

Further, another challenge of using RL to implement real-time adversarial attacks is the \emph{sparse rewards problem}, i.e., the agent only receives the reward at the end and it is difficult to obtain an estimation of the reward at each time point based on the observed data and past actions. For example, estimating the expected reward at a time point simply based on feeding the observed (partial) input at that time, superimposed with the corresponding perturbation, into the target model $f$ (if accessible) and using the classification confidence to calculate the reward will not yield reliable results, because the model’s prediction is not reliable when only partial input is given. In fact, although there have been many efforts to solve the sparse reward problem, many tasks still suffer from high computational overhead and training instability. However, for the adversarial example crafting problem, we could generate many trajectories of observation-action pairs using state-of-the-art non-real-time adversarial generation algorithms. This naturally leads us to use an \emph{imitation learning} and \emph{behavior cloning}~\cite{atkeson1997robot} strategy to overcome the sparse reward problem. We discuss it in the following section.

\squeezeups
\subsection{Imitation Learning Strategy}

Imitation learning is an RL technique that learns an optimal policy $\pi_{g}$ by imitating the behavior of an expert. Specifically, imitation learning requires a set of decision trajectories $\{\tau_1, \tau_2, ...\}$ generated by an expert, where each decision trajectory consists of a sequence of ``observation-action'' pairs, i.e., $\tau_i = \langle o_1^i, a_1^i, o_2^i, a_2^i, ..., o_n^i, a_n^i\rangle$. Such trajectories serve as demonstrations to teach the agent how to behave given an observation. We can extract all expert observation-action pairs from the trajectories and form a new dataset $\mathcal{D}=\{(o_1^1, a_1^1), (o_2^1, a_2^1), ..., (o_n^1, a_n^1), (o_1^2, a_1^2), (o_2^2, a_2^2), ...\}$. By treating $o$ as the input feature and $a$ as the output label, we could learn $\pi_g: a_{t}=g(o_t)$ in a supervised learning manner using traditional algorithms.

Specifically for the adversarial example crafting problem, we can use state-of-the-art non-real-time attack models to generate ``sample-perturbation'' pairs $\langle (\boldsymbol{x}^1, \boldsymbol{r}^1), (\boldsymbol{x}^2, \boldsymbol{r}^2), ... \rangle$ as decision trajectories by feeding different original samples $\boldsymbol{x}^i$ and collecting the corresponding output perturbations $\boldsymbol{r}^i$. Here, both $\boldsymbol{x}^i$ and $\boldsymbol{r}^i$ consist of a sequence of $x$ and $r$, using the definition of observation $o$ and action $a$ in Section~\ref{sec:formalism}. We can convert each $x$ and $r$ to $o$ and $a$, and then build a training set $\mathcal{D}$ and use supervised learning to learn $\pi_g$. 

\squeezeups
\subsubsection{Choice of Expert}
\label{sec:expert}

We use a state-of-the-art non-real-time adversarial example crafting technique as the expert. Over the last few years, many new attack techniques have been developed and shown to be effective. These techniques can be roughly classified into two categories. The first category includes gradient-based methods such as FGSM, DeepFool, and the method presented in~\cite{carlini2018audio}; these are typically based on deterministic optimization algorithms. The second category consists of gradient-free methods such as the methods presented in~\cite{alzantot2018did,su2019one}; these are typically based on stochastic optimization algorithms. Which method works better as an expert depends not only on the attack success rate; other important criteria include:

\textbf{1. Flexibility of adding additional constraints.}
There are two reasons why we prefer an expert that provides some flexibility of adding additional constraints besides making the perturbation imperceptible. First, we ultimately need to learn $\pi_g$ from the trajectories generated by the expert using some supervised learning method, which inevitably will contain some error. We can add some regularization on the trajectories (e.g., perturb only after a specific time point) to simplify the supervised learning task, which requires additional constraints on the expert. 
Second, in realistic attack scenarios, the attacker usually faces additional constraints, e.g., when an attacker attempts to fool a speech recognition system by playing the perturbation over the air using a speaker, the frequency range of the perturbation is subject to the characteristics of the speaker. In general, stochastic optimization algorithms are more flexible than deterministic optimization algorithms for adding additional complex constraints.

\textbf{2. Attacker's knowledge.} The attacker's knowledge required for the proposed real-time adversarial attack follows exactly the chosen expert policy. Hence, the attacker should choose the expert policy according to the attack scenario. 

\textbf{3. Determinism of the expert.}
While state-of-the-art adversarial example crafting approaches are highly effective in terms of success rate, there is no guarantee that the generated perturbation is globally optimal. Specifically, perturbations generated for the same input sample using a stochastic optimizing algorithm can vary with the random seed since the optimization solutions might stop at different sub-optimal points, which will make the mapping $o\mapsto a$ ill-defined and increase the difficulty of training $\pi_g$. Therefore, a deterministic expert is preferred.


\begin{figure}[t]
  \centering
  \includegraphics[width=7.0cm]{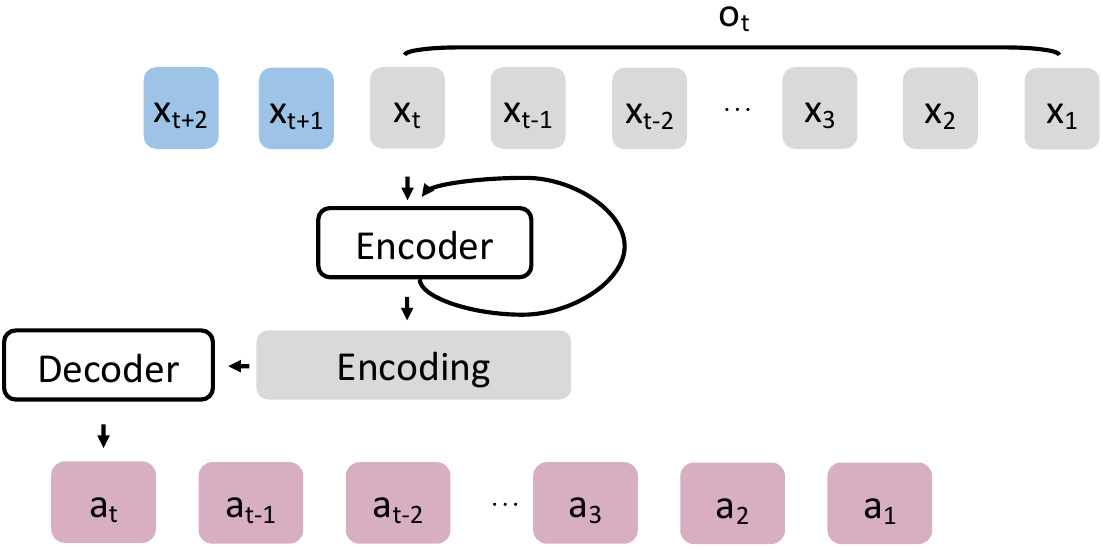}
  \caption{Illustration of the training process. Note that the output action only depends on the current observation $o_t$.}
  \label{fig:3}
  \squeezeup
  \squeezeup
\end{figure}

\subsubsection{Computational Overhead and Speed}
\label{sec:comp}

Existing adversarial example crafting techniques can be computationally expensive due to the complexity of optimization, e.g., the method in~\cite{carlini2018audio} requires about one hour to craft a single speech adversarial example. Stochastic optimization algorithms typically need to call the target model (or the substitute model) hundreds or thousands of times to find the solution. However, since we use a deep neural network $g$ to substitute optimization, no matter which expert we choose to imitate, the computational overhead for generating an adversarial perturbation for one time point is fixed to be the inference time of $g$ (denoted by $t_g$, which is the computational delay). In the real-time scenario, if the input sample frequency is higher than $\frac{1}{t_g}$, then the generator is not fast enough to catch up with the streaming input. The attacker then needs to lower the update frequency by modifying $g$ to do batch processing, i.e., generate a batch of $n_{batch}$ actions for $n_{batch}$ future points in one inference, which could lower the delay requirement by $n_{batch}$ times.

\begin{algorithm}[!t]
\caption{Real-time Adversarial Attack}\label{alg:1}
\begin{algorithmic}[1]
\Require{
\quad \newline
Original dataset $\mathcal{X}=\{\boldsymbol{x}^i\}$ where each $\boldsymbol{x}^i=\{x^i_t\}$ \newline
\textbf{Non-real-time} adversarial example generator (expert) $g_e$}

\Algphase{Phase 1: Generate Expert Demonstrations}
\textbf{Input:}  Original sample set $\mathcal{X}$ \newline
\textbf{Output:}  Expert decision trajectory set $\mathcal{D}$
\State initialize $\mathcal{D}$ as an empty set
\For {each $\boldsymbol{x}^i \in \mathcal{X}$}
\State $\boldsymbol{r}^i= \{r^i_t\} = g_e(\boldsymbol{x}^i)$
\State initialize trajectory $\tau_i$ as an empty set
\For {each time point $t$ of $\boldsymbol{x}^i$} 
\State $o^i_t=\{x^i_1, x^i_2, ... x^i_t\}$
\State $a^i_t=r_{t+d+1}^i$
\State add $(o^i_t, a^i_t)$ to $\tau_i$
\EndFor
\State add $\tau_i$ to $\mathcal{D}$
\EndFor 
\State \textbf{return} $\mathcal{D}$

\Algphase{Phase 2: Train Realtime Adversarial Example Generator}
\textbf{Input:} Expert decision trajectory set $\mathcal{D}$\newline
\textbf{Output:} Real-time adversarial example generator $g_r$
\State initialize $g_r$ as a recurrent network with parameter $\theta$

\For{each trajectory $\tau_i \in \mathcal{D}$}
\newline \Comment{maintain RNN states for each $t$ to expedite computing}
\For{each time point $t$} 
\State calculate the predicted action $\hat{a^{i}_t} = g_r(o^i_t)$
\State calculate the loss $l$ between the predicted action \newline 
\hspace*{2.8em} $\hat{a^i_t}$ and the expert's action $a^i_t$
\State update $\theta$ to minimize the loss $l$
\EndFor
\EndFor
\State \textbf{return} $g_r$

\Algphase{Phase 3: Conduct Real-time Adversarial Attack}
\textbf{Input:} Streaming observations $\boldsymbol{o}=\{o_1, o_2, ..., o_t, ...\}$ 
\State \textbf{at} each time point $t$ \textbf{do}
\State \hspace*{2.8em} $a_t = g_r(o_t)$
\State \hspace*{2.8em} execute action $a_t$
\end{algorithmic}
\end{algorithm}

\squeezeups
\subsection{Implementation}
\label{sec:imp}

Once we form the dataset $\mathcal{D}=\{(o_1, a_1), (o_2, a_2), ...\}$ consisting of observation-action pairs from the expert's decision trajectory, we form the real-time adversarial generator $g$ as a deep neural network and learn from the dataset. Note that each input $o$ is a sequence of variable length; so it is natural to use a recurrent neural network as part of the network. Specifically, the neural network can be divided into two parts: the encoder and the decoder. The encoder is a recurrent neural network that maps a variable length input into a fixed dimensional encoding. We expect that the learned encoding contains useful features from $o$; the decoder then makes the decision of the action, e.g., in the example in Figure~\ref{fig:ilus}, we expect that the encoding expresses which cluster the data sample belongs to, and the decoder can find the optimal perturbation based on this information. We can then calculate the error between the predicted action and the ground truth action and use standard back-propagation to update $g$.

Assume that we have $n_t$ trajectories and each trajectory consists of $n$ observation-action pairs. The dataset has $n\times n_t$ samples, which can be very large and will make the training slow. In fact, observations from the same trajectory are highly dependent, i.e., the only difference between $o_{t+1}$ and $o_{t}$ is that $o_{t+1}$ has one more observed point $x_{t+1}$; therefore there will be a lot of repetitive computation of the recurrent neural network (i.e., the encoder). In order to expedite the training, we should train observation-action pairs from the same trajectory in a batch, i.e., after obtaining $a_t$ from feeding input $o_t$ into $g$, we do not feed a new input $o_{t+1}$ into $g$. Instead, we feed $x_{t+1}$ into $g$ and obtain the output of $g$ as $a_{t+1}$. Figure~\ref{fig:3} illustrates this training process. Specifically, this approach avoids any repetitive encoder computation and can be viewed as a sequence to sequence training. Note that the predicted actions are only dependent on the current observation (i.e., they are not based on any future observations), which is different from standard sequence to sequence training used in other applications such as machine translation where the intermediate encoding contains information of the entire input sample. The pseudocode of the proposed algorithm is shown in Algorithm~\ref{alg:1}. 

It is worth mentioning that although in this paper, we focus on using the basic behavior cloning algorithm for simplicity, there are many more advanced algorithms (e.g., Dataset Aggregation~\cite{ross2011reduction}) in imitation learning and reinforcement learning that can further improve the attack performance, e.g., it is possible to design a remedy mechanism for the real-time adversarial perturbation generator that allows it to adjust its future strategy if it realizes it has previously made a wrong decision. Hence, formalizing the real-time attack into a reinforcement learning problem is not only natural, but also allows us to apply existing tools and algorithms.

\section{Case Study: Attacking a Voice Command Recognition System}

In the previous section, we introduced the general real-time adversarial attack framework in a relatively abstract way; in this section, we further show how to adopt the framework in a realistic task: the audio adversarial attack\footnote{Code and demos are available at \url{https://github.com/YuanGongND/realtime-adversarial-attack}}.

\subsection{Target Model and Attack Scenario}
\label{sec:target_model}

The goal is to attack a voice command recognition system based on a convolutional neural network~\cite{sainath2015convolutional}. This model is used as an official example for Tensorflow\footnote{\url{www.tensorflow.org/tutorials/sequences/audio_recognition}}, it is easy to reproduce, and has also been used as the target model for attacks in~\cite{alzantot2018did}. We train the voice command recognition model exactly as in the implementation of the Tensorflow example using the voice command dataset~\cite{warden2018speech}, except that we only use 80\% of the data for training, allowing us to use the other 20\% for testing. Most audio samples are of exact 1-second length with a sampling rate of 16 kHz; all other samples are padded to be also of 1 second for consistency. The model can classify ten keywords: ``yes'', ``no'', ``up'', ``down'', ``left'', ``right'', ``on'', ``off'', ``stop'', and ``go''. The trained model achieves 88.7\% accuracy on the validation set. 

\begin{figure}[t]
  \centering
  \includegraphics[width=8.2cm]{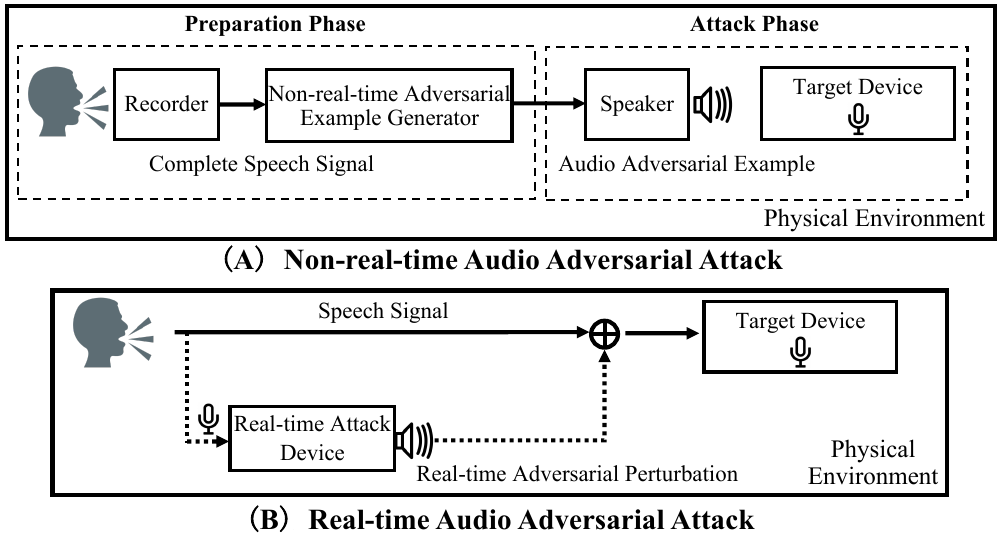}
  \caption{Illustration of the non-real-time (upper figure) and real-time (lower figure) audio adversarial attack.}
  \label{fig:audio_scene_ilus}
  \squeezeup
\end{figure}

The proposed real-time scheme can greatly increase the real-world threat of the audio adversarial attack. As illustrated in Figure~\ref{fig:audio_scene_ilus}, compared to previous non-real-time audio adversarial attack technologies presented in~\cite{carlini2016hidden,yakura2018robust,gong2018overview,qin2019imperceptible}, the key advantage of the real-time audio adversarial attack scheme is that only by using this scheme the attacker is able to conduct attacks to an \textbf{on-going} session, i.e., an on-going human-computer interaction, and interfere with the voice command currently being spoken by a human speaker. This is because previous non-real-time adversarial attack approaches needed a ``preparation stage'', where the attacker obtains a \textbf{complete} original speech sample, designs specific adversarial perturbations for this sample, and adds a perturbation to the original sample to build a malicious adversarial example. Then, in the ``attack phase'', the attacker needs to initialize a new session with the target system and then replay the prepared malicious adversarial example. The application of such an attack is relatively limited, because during the attack phase, if the user is near the target system, then no matter how close the malicious sample sounds to a benign sample, it will be suspicious to the user; if the user is not near the target system, then it is not necessary to make the malicious sample imperceptible to humans. Further, it is not always easy or even possible to initiate a new session in security-sensitive systems. In contrast, the real-time adversarial attack scheme does not need a preparation phase; instead, it continuously processes the speech spoken by the user and emits the adversarial perturbation, which is superimposed with the original signal over the air in a real-time manner. In practice, the attack can be implemented by placing a device (e.g., a smartphone) equipped with a microphone and a speaker and installed with the real-time attack software near the target device. 

\squeezeups
\subsection{Adversarial Attack Settings}
\label{constraints}
We perform the non-targeted attack in a semi-black box setting, i.e., we assume that the attacker can call the target model an unlimited number of times and get the corresponding predictions and confidence score, but has no knowledge about the model details (architectures, algorithm, and parameters). It is a realistic setting for speech recognition system attacks, because the loss function of many speech recognition models cannot be differentiable with respect to the input, and most state-of-the-art systems are cloud-based, which makes it difficult to obtain full knowledge of the model and perform a white-box attack. For example, the front end of our target model is not a neural network, but a set of filter banks extracting Mel-frequency cepstrum features, so it is hard to calculate the gradient of the loss function with respect to the input waveform, even when we have a copy of the model~\cite{alzantot2018did}; Google Speech is a commercial cloud-based model which is hard for the attacker to obtain full knowledge about its design. However, it allows users to upload speech samples and freely obtain predictions and confidences scores, which provides opportunities for semi-black box attacks.

\begin{figure}[h]
  \centering
  \includegraphics[width=8.05cm]{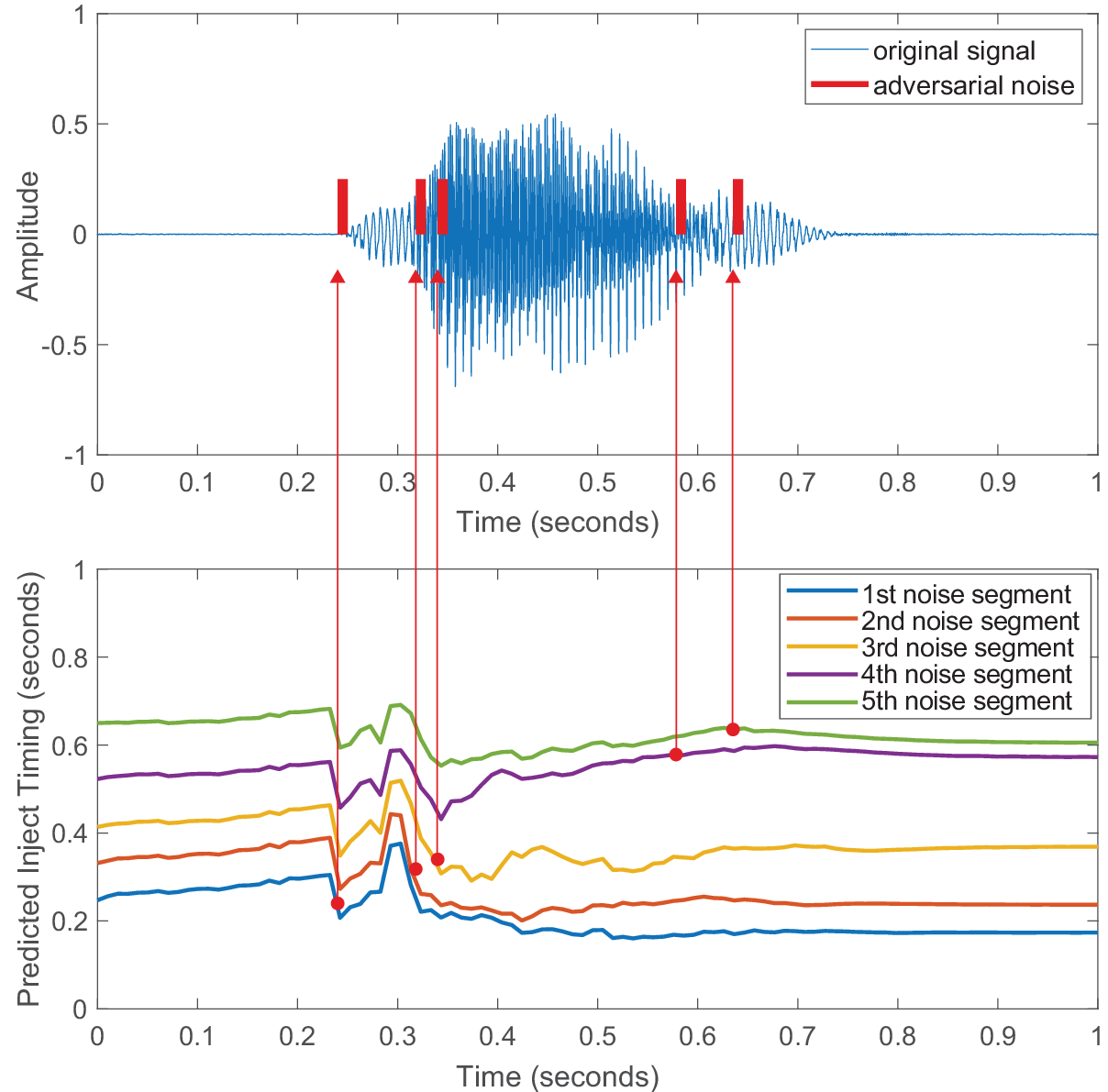}
  \caption{Illustration of the proposed real-time audio adversarial attack using a real sample. The real-time perturbation generator continuously predicts the best timing to emit each of the five 0.01-second adversarial noise segments based on the observation, and conducts emission immediately once the predicted timing is equal to or earlier than the current time point. We can observe that the prediction changes dramatically when the speech signal is observed, but barely changes when the silent period is observed, indicating that the generator makes decisions mainly based on the informative part of the signal, and is able to correct them given more observation. The actual emission time points are shown with red dots in the lower figure; they are unlikely to be the optimal choice with full observation, but are the best guess at that time given partial observation.}
  \label{fig:inject_ilus}
  \squeezeup
\end{figure}

In order to emulate a realistic situation, in this example, we apply the following 
constraints to the adversarial perturbation. First, we constrain the $l_0$ norm of the adversarial perturbation, i.e., we limit the number of non-zero points of the perturbation. This is because limiting the $l_1$ or $l_2$ norm will make the amplitude of the noise small and does not pose an over-the-air threat; so it is more reasonable to generate short, but relatively loud perturbations. Second, we require that the non-zero points of the perturbations must form clusters as consecutive noise segments. This is because it is impossible for an electronic speaker to generate a signal of a few non-consecutive non-zero points due to the limitation of its dynamic characteristics. In this sample, we perturb five 0.01-second segments and, for simplicity, the scales of the points in one segment are fixed and identical (i.e., the noise frequency is an integral multiple of the sampling frequency), but each noise segment can be any physically realizable signal the attack desires. Data points that have amplitudes over 1 are clipped to 1. These two constraints also greatly lower the computational complexity for the real-time adversarial perturbation generator, which now only needs to decide the timing of emitting each of the five noise segments. In this sample, we focus on the decision-making process, so we do not consider the signal attenuation and distortion during transmission through the air. An illustration of the proposed adversarial perturbation is shown in the upper part of Figure~\ref{fig:inject_ilus}.

\squeezeup
\subsection{The Expert}
\label{sec:speech_expert}

Since we are performing a semi-black box attack, and to ensure realism, we add non-standard constraints to the optimization problem. Following the discussion in Section~\ref{sec:expert}, we choose a stochastic optimization based adversarial example crafting technique as the expert. Specifically, we take the \emph{differential evolution} optimization~\cite{storn1997differential}, which was previously used for the ``one-pixel'' attack~\cite{su2019one} on image recognition systems with similar constraints to our proposed attack. We extend it for use as audio attacks, and then use it as the expert. In our case, the candidate solution of the optimization is a 5-tuple consisting of the starting points of each noise segment (sorted). The optimization objective is to minimize the confidence score of the original label. At each iteration, the fitness of each candidate solution is calculated and new candidate solutions are produced using the standard differential evolution formula.

\begin{figure}[t]
  \centering
  \includegraphics[width=8.3cm]{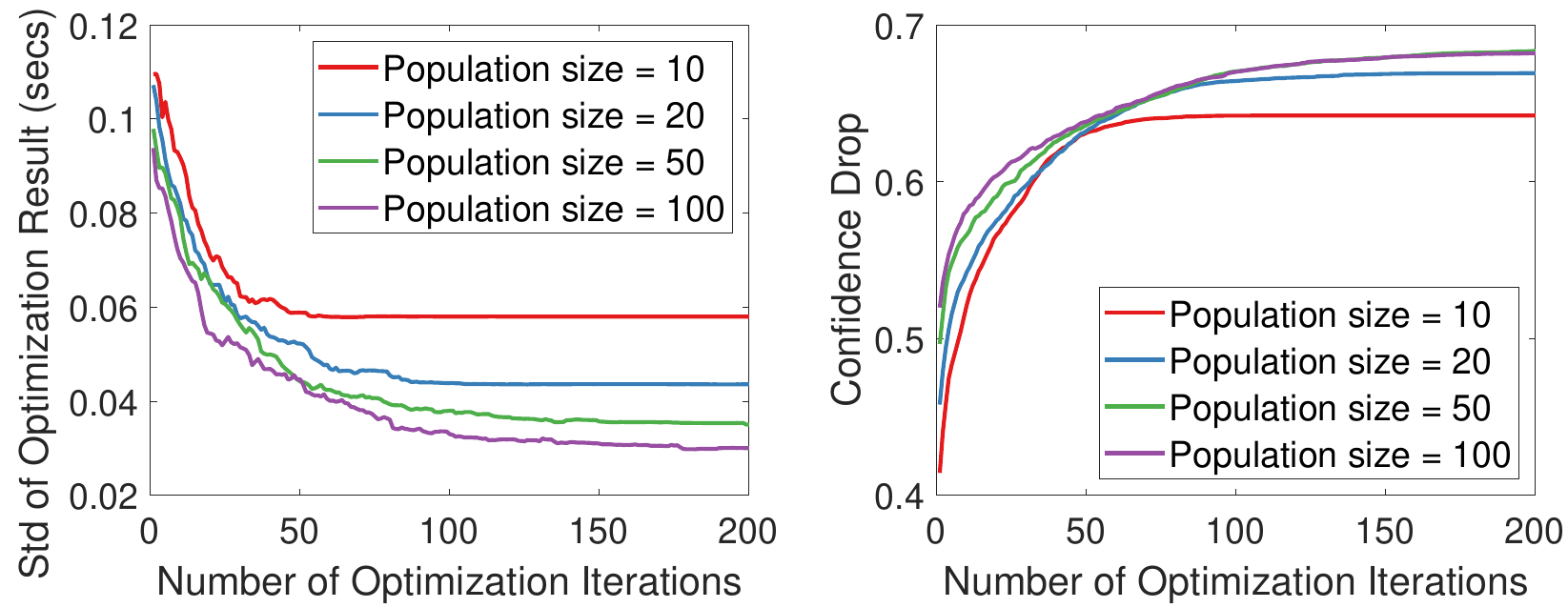}
  \caption{The standard derivation of the optimization result using different random seeds (left figure) and the confidence score drop of the original class led by the expert attack (right figure) with different numbers for optimization iterations and population size.
}
  \label{fig:expert_performance}
  \squeezeup
  \squeezeup
\end{figure}

The differential evolution algorithm has two main parameters: the population size and the number of iterations. On one hand, we want the optimization result to be optimal and deterministic (i.e., the result is invariant to random seeds), which requires large parameters. On the other hand, the computational overhead is linearly proportional to the population and the iteration number, and evaluating the fitness of each candidate solution requires calling the DNN based target model once. Therefore, in order to generate the dataset consisting of over 20,000 trajectories for imitation learning over a reasonable time, we have to limit the population and the iteration number. As shown in Figure~\ref{fig:expert_performance}, we test the performance and the standard derivation of the optimization result with different random seeds. We find that population size = 10 and iteration number = 75 provide a good balance between performance and computational overheads and use these values in our experiments. For each audio in the training set, we use the expert to generate a perturbation in the form of a 5-tuple. Note that each audio consists of 16,000 observations, and thus forms 16,000 observation-tuple pairs (a decision trajectory), where the tuple is identical for all observations since the optimal perturbation does not change with the observation.

\subsection{Training the Real-time Adversarial Perturbation Generator}
\label{sec:audio_train}

\subsubsection{Input and Output of the Network}

The real-time adversarial perturbation generator is implemented using a deep neural network; the input of the network is simply an observation $o$ (of variable length), the output of the network is a 5-tuple of the same definition as the solution of the differential evolution optimization algorithm, i.e., 5 time points to emit noise segments. The tuple can be easily converted to action $a$ using the following rule: if the current estimated best emission timing is equal to or earlier than the current time point, then immediately emit the noise.

\subsubsection{Batch Processing}

The frequency of the speech signal (i.e., 16 kHz) is much higher than the possible update speed of the real-time adversarial perturbation generator. Therefore, we apply batch processing as mentioned in Section~\ref{sec:comp}; specifically, the adversarial generator updates every 0.01 second and each update makes a decision on the actions for 0.01 seconds, so the delay is also 0.01 seconds. Note that while the update period and noise segment length are identical, they are not related.

\subsubsection{The Network Architecture}

As shown in Table~\ref{tab:arc}, we use an end-to-end neural network. Since the input is an observation of variable length $t$, as a standard signal processing technique, we cut it into $\lceil t/160\rceil$ frames,  where 160 is the frame length. We then use a series of convolution and pooling layers to extract the features. The features of each frame are then sequentially fed into the long short-term memory (LSTM)~\cite{hochreiter1997long} layers to obtain the encoding, and two dense layers decode the encoding as the output. This basically follows the architecture shown in Figure~\ref{fig:3}: the layers before the LSTM layers are the encoder, and those after the LSTM layers are the decoder. We use 1e-3 as the learning rate, mean square loss, and ADAM optimizer~\cite{kingma2014adam} for training. We train data samples in the same trajectory in a batch to expedite the computations as discussed in Section~\ref{sec:imp}.

\begin{table}[]
\centering
\scriptsize
\begin{tabular}{|c|c|}
\hline
\textbf{Layer Name} & \textbf{Output Dimension} \\ \hline
Input               & (t, 1)                    \\ \hline
Framing             & ($\lceil t/160 \rceil$, 160)        \\ \hline
Conv1 / Pooling     & ($\lceil t/160\rceil$, 80, 16)     \\ \hline
Conv2 / Pooling     & ($\lceil t/160\rceil$, 40, 32)     \\ \hline
Conv3 / Pooling     & ($\lceil t/160\rceil$, 20, 48)     \\ \hline
Conv4 / Pooling     & ($\lceil t/160\rceil$, 10, 64)     \\ \hline
Flatten             & ($\lceil t/160\rceil$, 640)        \\ \hline
LSTM * 3            & (256)                     \\ \hline
Dense 1             & (256)                     \\ \hline
Dense 2             & (128)                     \\ \hline
Output              & (5)                      \\ \hline
\end{tabular}
\caption{The network details and the output dimension of each layer.}
\label{tab:arc}
\squeezeup
\end{table}

\squeezeups
\subsection{Experiments}
\label{sec:result}

In our experiments, we test the dataset and target model mentioned in Section~\ref{sec:target_model}. The data is split as follows: we first hold out 20\% of the data as the test set (test set 2) for evaluating the attack performance; so it is not seen by the target model and the attack model. We use the other 80\% of the data to train the target voice recognition model; this same set is then reused to develop the attack model.  Specifically, we use 75\% of this set to train the attack model (attack training set), 6.25\% for validation, and 18.75\% for testing (test set 1). Therefore, test set 1 is seen by the target model, but not seen by the attack model. We then generate the expert demonstration of optimal emission timing using the method mentioned in Section~\ref{sec:speech_expert} for each speech sample in the attack train set. Since in our setting, the amplitude of each noise segment is a given fixed value, it is expected (and is proven by our experiments later) that the expert demonstration of the optimal emission timing varies with the given amplitude value because the emission strategy may be different for different noise amplitude. In this experiment, we generate two versions of expert demonstrations using noise amplitude of 0.1 and 0.5, respectively. Note that although the expert demonstration of optimal emission time points are optimized based on a given noise amplitude, the attacker can emit noise of any amplitude as desired at these time points in the test phase, which might lead to a sub-optimal attack performance. We discuss it in detail in the next section.

We then train the real-time adversarial perturbation generator to learn from the expert demonstrations using the approach described in Section~\ref{sec:audio_train}. We use two metrics to evaluate the attack performance: 1) attack success rate (in the non-targeted attack setting, success means that the prediction of the perturbed sample is different from that of the original sample) and 2) confidence score drop of the original class led by the attack (which measures the confidence of the attack).

\begin{figure}[t]
  \centering
  \includegraphics[width=8.3cm]{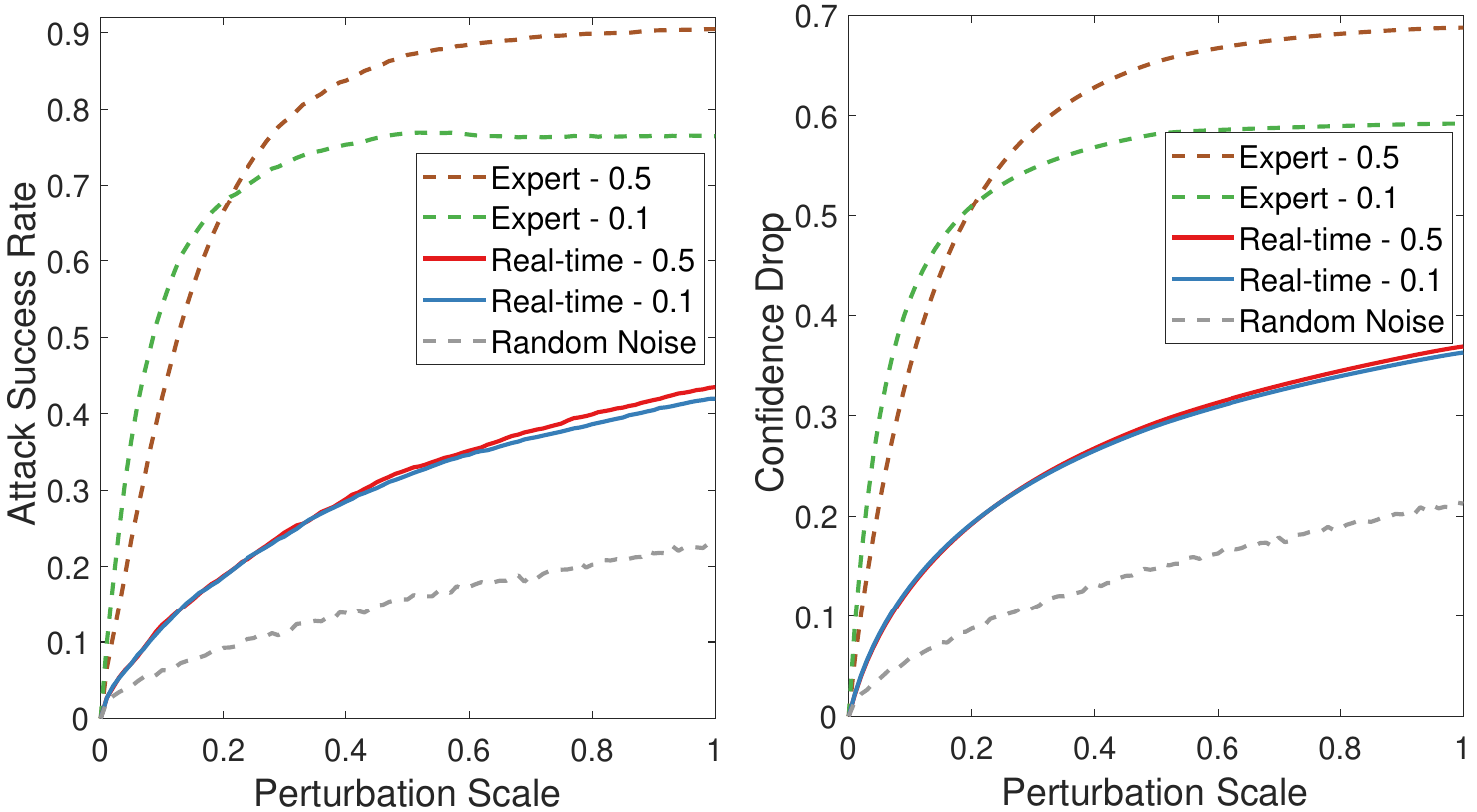}
  \caption{The attack successful rate and the confidence score drop led by the attack with different perturbation scale.}
  \label{fig:overall_result}
\end{figure}

\begin{figure}[t]
  \centering
  \includegraphics[width=8.3cm]{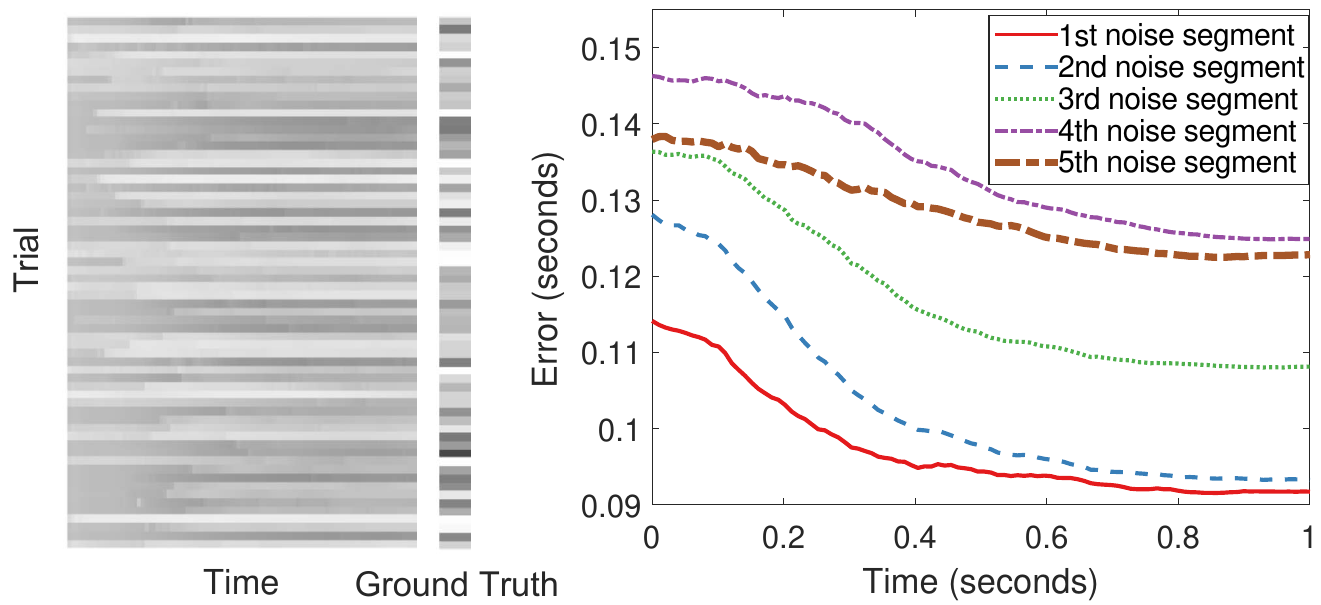}
  \caption{Left: the adversarial perturbation generator's estimate on the optimal noise emission timing (shown by color; dark represents later time, light represents earlier time) of the first noise segment at each time point. Each row represents one attack trial and 64 trials are shown in total. Right: the mean prediction error over time.}
  \label{fig:trials}
  \squeezeup
  \squeezeup
\end{figure}

\subsubsection{Overall Result}

We show the attack performance on test set 1 of two non-real-time experts (optimized for perturbation amplitude of 0.1 and 0.5, respectively) and corresponding learned real-time adversarial perturbation generators in Figure~\ref{fig:overall_result}. We observe that:

First, the attack success rate of the adversarial perturbation generator is up to \textbf{43.5\%} (when perturbation amplitude is 1), which is about half of the best non-real-time expert (90.5\%) and clearly outperforms the random noise. For most perturbation amplitudes, the attack success rate of the real-time attack is 30\%-50\% of that of the expert.
    
Second, the attack performance of the expert varies with the perturbation amplitude it is optimized for. It is not surprising that the expert optimized for small noise amplitude of 0.1 performs better when the actual emission amplitude is small ($<$ 0.23) while the expert optimized for large amplitude of 0.5 performs better when the actual emission amplitude is large ($\geq$ 0.23). This difference also shows in the corresponding real-time adversarial perturbation generators, but the impact is much smaller, which gives the attacker a nice property that the attack performance does not drop much when the actual and expected noise amplitude are different (e.g., for audio adversarial attacks, it is hard for the attacker to know the actual amplitude of the noise signal received by the target system due to signal attenuation, but it does not matter).
    
Third, we further conduct the same test on the test set 2 (attack success rate up to 42.2\%) and have not found a substantial difference between the result of test set 1 and 2, indicating the attack model can be generalized to data samples that have not been seen by the target model.

\subsubsection{Real-time Dynamics}

We next discuss how the proposed adversarial perturbation generator works in a real-time manner; towards this end, we plot the dynamics of 64 attack trials for 64 different input samples in the left part of Figure~\ref{fig:trials}. Each row represents one attack trial, which shows the adversarial perturbation generator's estimate on the optimal emission timing of the first noise segment at each time point. We place the ground truth on the right for reference. At the beginning of each attack, when no data is observed yet, the adversarial perturbation generator outputs a prior guess which has similar values for different samples, but with more data observed, the estimate gradually improves and finally approaches the ground truth. We can also observe that the amount of observations needed for correct estimates differs among the trials. This is because the voice command samples have different lengths of silence periods at the beginning, which does not contain information helpful to predict an adversarial perturbation. This can be further verified by the detailed dynamics of a real sample shown in Figure~\ref{fig:inject_ilus}, where we can find that the estimation of the adversarial perturbation generator changes dramatically when the speech signal is observed, but barely changes when the silent period is observed, indicating the generator makes decisions mainly based on the informative part of the signal, and is able to correct them given more observation. In this sample, the estimation does not become stable until half of the speech signal is observed, but three noise segments are already emitted by this time point, showing the trade-off between the observation and action space, i.e., the attacker needs to emit the adversarial noise immediately when the current best-guess timing with partial observation arrives, otherwise the timing will pass and the emission cannot be implemented. We also show the mean absolute prediction error over time in the right part of Figure~\ref{fig:trials}, which demonstrates that the adversarial generator indeed improves with more observations.

\begin{figure}[t]
  \centering
  \includegraphics[width=6.3cm]{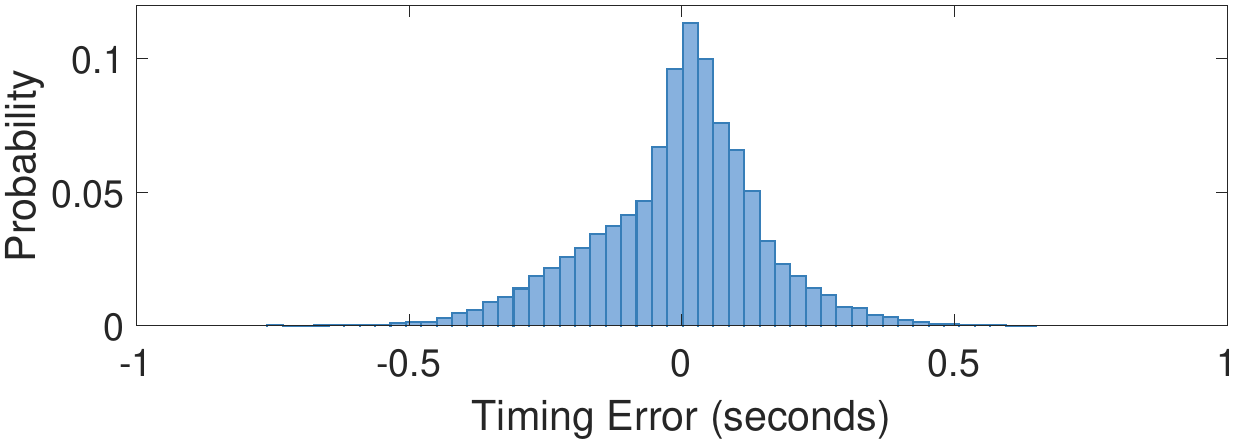}
  \caption{The distribution of the actual emission timing errors.}
  \label{fig:inject_error}
  \squeezeup
  \squeezeup
\end{figure}

\subsubsection{Error Analysis}

Finally, we analyze the error of the real-time adversarial generator. There are two main types of errors causing the performance gap between the expert and the real-time adversarial generator: prediction error and real-time decision error. The proposed real-time generator essentially tries to build the mapping between the (partial) input and output of the differential evolution optimization, while this substantially speeds up the computing, it is challenging to learn such a mapping. Specifically, in our setting, the output of the stochastic optimization algorithm adopted by the expert is not deterministic (shown in the left part of Figure~\ref{fig:expert_performance}), which makes learning such a mapping even harder. As shown in the right part of Figure~\ref{fig:trials} and, even after the real-time generator observes the full data, its prediction still has a certain amount of prediction errors. Further, as discussed in the previous section, the real-time adversarial generator may emit noise segments when it does not have a reliable estimation due to the observation-action space tradeoff. We show that the distribution of the actual timing error in Figure~\ref{fig:inject_error}, which obeys a zero-centered bell-shaped distribution, and the errors of most trials are small. Statistically, the mean actual timing error (i.e., the difference between the actual emission time point and the expert's demonstration) is 0.1135 seconds, which is slightly larger than the prediction error (i.e., the difference between the predicted emission time point with full observation and the expert's demonstration) of 0.1091 seconds. This indicates that the main error of our attack model is the prediction error, which can be improved by further reducing the instability of the expert and optimizing the deep neural network architecture. The proposed adversarial perturbation is audible even when the amplitude is small. It sounds similar to ``usual'' noise experienced by electronic speakers (e.g., buzzing, interference, etc.), which makes the perturbation appear not suspicious.

\section{Conclusions and Future Work}

In this work, we propose the concept of real-time adversarial attacks and show how to attack a streaming-based machine learning model by designing a real-time perturbation generator that continuously uses observed data to design optimal perturbations for unobserved data. We use imitation learning and behavioral cloning algorithm to train the real-time adversarial perturbation generator through the demonstrations of a state-of-the-art non-real-time adversarial perturbation generator. The case study (voice command recognition) and results demonstrate the effectiveness of the proposed approach. Nevertheless, we observe a certain performance gap between the real-time and the non-real-time adversarial attack when the basic behavior cloning algorithm is used. In our future research, we plan to study how to adopt more advanced reinforcement learning tools to improve the performance of decision making process,  e.g., when the real-time adversarial perturbation generator realizes it has previously made a wrong decision, could it adjust its future strategy to make it up? On the other hand, we plan to study the defense strategy to protect real-time systems against such real-time adversarial attack.

\newpage

\bibliographystyle{named}
\bibliography{ijcai19}

\end{document}